\documentclass[aps,twocolumn,showpacs,superscriptaddress,amsmath,amssymb,nofootinbib]{revtex4-1}

\usepackage{graphicx}
\usepackage{dcolumn}
\usepackage{bm}
\usepackage{color}
\usepackage{txfonts}
\usepackage{hyperref}
\usepackage{soul}

\begin{document}
\title{Efficiency at the maximum power output for simple two-level heat engine}

\author{Sang Hoon Lee}
\affiliation{School of Physics, Korea Institute for Advanced Study, Seoul 02455, Korea}

\author{Jaegon Um}
\affiliation{Quantum Universe Center, Korea Institute for Advanced Study, Seoul 02455, Korea}
\affiliation{CCSS, CTP and Department of Physics and Astronomy, Seoul National University, Seoul 08826, Korea}

\author{Hyunggyu Park}
\affiliation{School of Physics, Korea Institute for Advanced Study, Seoul 02455, Korea}
\affiliation{Quantum Universe Center, Korea Institute for Advanced Study, Seoul 02455, Korea}


\begin{abstract}
We introduce a simple two-level heat engine to study the efficiency in the condition of the maximum power output, depending on the energy levels from which the net work is extracted. In contrast to the quasi-statically operated Carnot engine whose efficiency reaches the theoretical maximum, recent research on more realistic engines operated in finite time has revealed other classes of efficiency such as the Curzon-Ahlborn efficiency maximizing the power output. We investigate yet another side with our heat engine model, which consists of pure relaxation and net work extraction processes from the population difference caused by different transition rates. Due to the nature of our model, the time-dependent part is completely decoupled from the other terms in the generated work. We derive analytically the optimal condition for transition rates maximizing the generated power output and discuss its implication on general premise of realistic heat engines. In particular, the optimal engine efficiency of our model is different from the Curzon-Ahlborn efficiency, although they share the universal linear and quadratic coefficients at the near-equilibrium limit. We further confirm our results by taking an alternative approach in terms of the entropy production at hot and cold reservoirs.
\end{abstract}



\maketitle


\section{Introduction}
\label{sec:introduction}

The efficiency of heat engines is a celebrated topic of classical thermodynamics~\cite{HuangBook}. In particular, an elegant formula expressed only by hot and cold reservoir temperatures for the ideal quasi-static and reversible engine coined by Sadi Carnot has been an everlasting textbook example~\cite{Carnot1824}. That ideal engine, however, is not the most {\em efficient} engine any more when we consider its power (the extracted work per unit time), which has added different types of optimal engine efficiency such as the Curzon-Ahlborn efficiency for some cases~\cite{Chambadal1957,Novikov1958,Curzon1975}. Following such steps, researchers have taken simple systems to investigate various theoretical aspects of underlying principles of macroscopic thermodynamic engine efficiency in details~\cite{Hoppenau2013,Proesmans2015,Um2015,Holubec2015,JMPark2016,Ryabov2016,Shiraishi2016} and its microscopic fluctuation~\cite{Verley2014a,Verley2014b,Gingrich2014,Proesmans2015a,Proesmans2015b,Polettini2015}.

In this paper, we introduce a simple two-level heat engine model to explore the condition for the maximum power. In our model, the time-dependent part is completely decoupled from the rest of the formulation, which makes the analysis considerably simpler. We derive analytically a parameter relation between transition rates at the maximum power for a given temperature ratio. We compare the functional form of the optimal efficiency at the maximum power to previously known forms for some other cases. Our result 
shows a difference from the Curzon-Ahlborn efficiency, but shares the same asymptotic behavior up to the second order in a small
efficiency limit. We also take an alternative approach considering the entropy production at the reservoirs, and discuss its implication. Generalization to multi-level engines is considered, but a decoupling of the operating time does not happen, which
makes the analytic investigation quite complex.

\section{Two-level heat engine}
\label{sec:model}

\begin{figure*}
\includegraphics[width=0.9\textwidth]{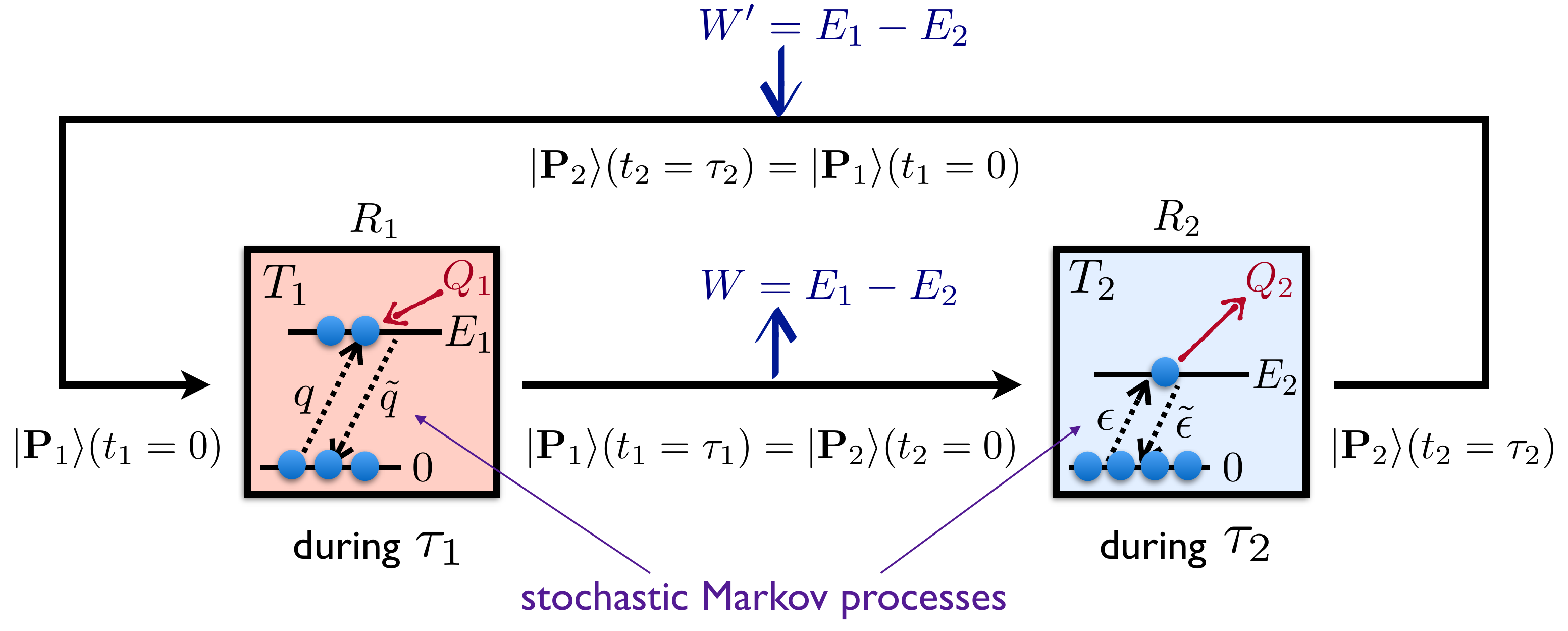}
\caption{Schematic illustration of our simple two-level heat engine, composed of two energy levels coupled with two heat reservoirs $R_1$ and $R_2$.}
\label{fig:schematic}
\end{figure*}

Figure~\ref{fig:schematic} illustrates our model. The two-level system is characterized by two discrete energy states composed of the ground state ($E = 0$) and the excited state ($E = E_1$ or $E = E_2$, depending on the reservoir of consideration). The transition rates from the ground state to the excited state are denoted by $q$ and $\epsilon$, respectively, and their reverse processes by $\tilde{q}$ and $\tilde{\epsilon}$. We assume $E_1 > E_2$ and $T_1 > T_2$. The system is attached to two different reservoirs: $R_1$ with temperature $T_1$ during time $\tau_1$, and $R_2$ with temperature $T_2$ during time $\tau_2$, and the adiabatic work extraction occurs in between. Although the amount of energy unit involving the work exchange is the same $(W = E_1 - E_2 = W')$ in Fig.~\ref{fig:schematic}, the net positive work is achievable due to the difference in the population of the excited states at the end of contact with $R_1$ and $R_2$, which is determined by model parameters as presented in Sec.~\ref{sec:max_power}.

\section{Engine efficiency}
\label{sec:max_power}

\subsection{Efficiency as a function of model parameters}
\label{sec:model_parameters}

The transition rates from the ground state to the excited state at reservoirs $R_1$ and $R_2$ are given as the following Arrhenius form,
\begin{equation}
\begin{aligned}
q/\tilde{q} = e^{-E_1/T_1} \,, \\
\epsilon/\tilde{\epsilon} = e^{-E_2/T_2} \,,
\end{aligned}
\label{eq:q_and_epsilon}
\end{equation}
respectively (we let the Boltzmann constant $k_B = 1$ for notational convenience), thus the inequality $0 < \epsilon < q < 1/2$ holds ($\epsilon < q$ is essential to get the positive amount of net work).
The average amount of work extracted from the system at $R_1 \to R_2$ and that given to the system at $R_2 \to R_1$ considering the population difference are given by
\begin{equation}
\begin{aligned}
\langle W \rangle = (E_1 - E_2) P_{1e} \,, \\
\langle W' \rangle = (E_1 - E_2) P_{2e} \,,
\end{aligned}
\label{eq:W_and_W_prime}
\end{equation}
respectively,
and $|\mathbf{P}_1 \rangle = \left( P_{1e} , P_{1g} \right)^T$ and $| \mathbf{P}_2 \rangle = \left( P_{2e} , P_{2g} \right)^T$ are the column vectors whose components represent the populations of excited and ground states (in that order) at the end of the contact with $R_1$ and $R_2$, respectively.
We then take the normalization convention $P_{1e} + P_{1g} = P_{2e} + P_{2g} = q + \tilde{q} = \epsilon + \tilde{\epsilon} = 1$, expressing the conservation of total population.
For notational convenience, we define the function $X(T,r)$ of temperature $T$ and transition rate $r$ as
\begin{equation}
X (T,r) \equiv T \ln(\tilde{r}/r) \,.
\label{eq:definition_X}
\end{equation}
Then, the average amount of heat to the system from $R_1$ and that from the system to $R_2$ are
\begin{equation}
\begin{aligned}
\langle Q_1 \rangle & = (P_{1e} - P_{2e}) X (T_1,q)  \\
\langle Q_2 \rangle & = (P_{1e} - P_{2e}) X (T_2,\epsilon) \,,
\end{aligned}
\label{eq:Q1_and_Q2}
\end{equation}
respectively,
based on the Schnakenberg entropy production for stochastic processes~\cite{Schnakenberg1976,Andrieux2004,Seifert2005}.
The average total entropy production during one cycle is given by the entropy change of the reservoir,
\begin{equation}
\begin{aligned}
\langle \Delta S \rangle & = - \frac{\langle Q_1 \rangle}{T_1} + \frac{\langle Q_2 \rangle}{T_2} \\
& = (P_{1e} - P_{2e}) \left[ \ln(\tilde{\epsilon}/\epsilon)- \ln(\tilde{q}/q)\right] \,.
\end{aligned}
\label{eq:entropy_change_one_cycle}
\end{equation}
Eqs.~\eqref{eq:W_and_W_prime} and \eqref{eq:Q1_and_Q2} ensure the energy conservation or the first law of thermodynamics
$\langle W \rangle - \langle W' \rangle = \langle Q_1 \rangle - \langle Q_2 \rangle$, considering Eq.~\eqref{eq:q_and_epsilon}.
The average net work extracted from the system is
\begin{equation}
\langle W_\mathrm{net} \rangle = \langle W \rangle - \langle W' \rangle = (P_{1e} - P_{2e}) \left[ X(T_1,q) - X(T_2,\epsilon) \right] \,,
\label{eq:Wnet}
\end{equation}
and the efficiency is given by the ratio
\begin{equation}
\eta = \frac{\langle W_\mathrm{net} \rangle}{\langle Q_1 \rangle} = 1 - \frac{X(T_2,\epsilon)}{X(T_1,q)} \,,
\label{eq:eta}
\end{equation}
independent of $\tau_1$ and $\tau_2$, and $\eta$ approaches $\eta_C = 1 - T_2/T_1$ (the Carnot efficiency~\cite{HuangBook,Carnot1824}) when $\epsilon \simeq q$, and meaningful only for $q > \epsilon$,
or $\langle W_\mathrm{net} \rangle > 0$.

\begin{figure*}
\begin{tabular}{ll}
(a) & (b) \\
\includegraphics[width=0.7\columnwidth]{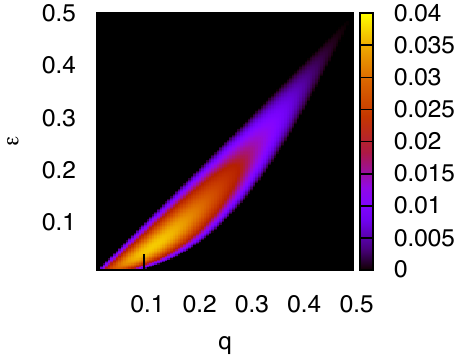} &
\includegraphics[width=0.7\columnwidth]{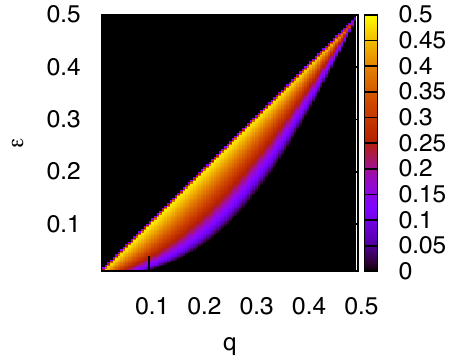} \\
(c) & (d) \\
\includegraphics[width=0.7\columnwidth]{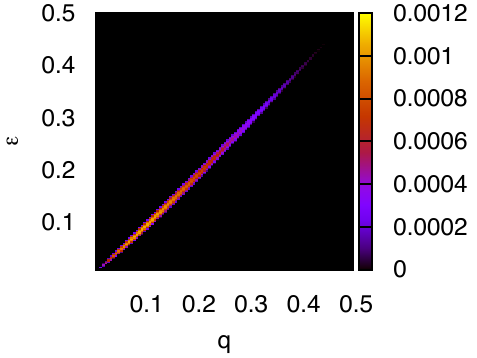} &
\includegraphics[width=0.7\columnwidth]{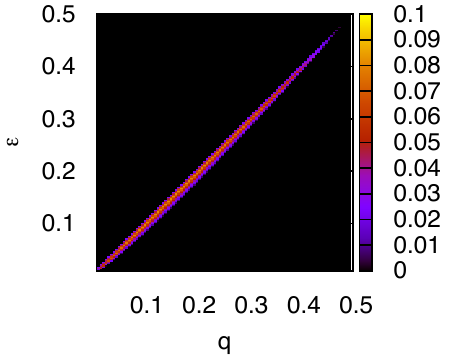} \\
\end{tabular}
\caption{(a) The average net work $\lim_{\tau \to \infty} \langle W_\mathrm{net} \rangle$ and (b) efficiency $\eta$ for $T_1 = 1$ and $T_2 = 1/2$,
and (c) $\langle W_\mathrm{net} \rangle (\tau \to \infty)$ and (d) $\eta$  for $T_1 = 1$ and $T_2 = 9/10$. For better visibility focused on the $\langle W_\mathrm{net} \rangle \ge 0$ regime, we set all of the negative values as $0$.}
\label{fig:Wnet_and_eta_3D}
\end{figure*}

Now let us consider the explicit form of populations at the excited states at end of each reservoir contact process, whose time evolution is given by the following linear differential equation system for given $q$ and $\epsilon$ values,
\begin{equation}
\begin{aligned}
\frac{d|\mathbf{P}_1\rangle}{dt_1} & =
\begin{bmatrix}
-\tilde{q} & q \\
\tilde{q} & -q \end{bmatrix}
|\mathbf{P}_1\rangle \,, \\
\frac{d|\mathbf{P}_2\rangle}{dt_2} & =
\begin{bmatrix}
-\tilde{\epsilon} & \epsilon \\
\tilde{\epsilon} & -\epsilon \end{bmatrix}
|\mathbf{P}_2\rangle \,,
\end{aligned}
\end{equation}
where $0 \le t_1 \le \tau_1$ and $0 \le t_2 \le \tau_2$ are the intermediate time spent in contact with $R_1$ and $R_2$, respectively.
As the populations do not change during the adiabatic work extraction (supply) processes, we get the {\em circular} boundary
condition as
$P_{1e} (t_1 = 0,t_2 = \tau_2) = P_{2e} (t_1 = \tau_1,t_2 = \tau_2)$ and $P_{2e} (t_1 = \tau_1,t_2=0) = P_{1e} (t_1 = \tau_1,t_2 = \tau_2)$. Thus, the solution at $t_1 = \tau_1$ and $t_2 = \tau_2$ is given by
\begin{equation}
\begin{aligned}
P_{1e} & = \displaystyle \frac{q(1-e^{-\tau_1})+\epsilon(1-e^{-\tau_2}) e^{-\tau_1}}{1-e^{-(\tau_1+\tau_2)}} \,, \\
P_{2e} & = \displaystyle \frac{\epsilon(1-e^{-\tau_2})+q(1-e^{-\tau_1})e^{-\tau_2}}{1-e^{-(\tau_1+\tau_2)}} \,,
\end{aligned}
\label{eq:P1_and_P2_solution}
\end{equation}
and $\lim_{\tau_1, \tau_2 \to \infty} P_{1e} = q$ and $\lim_{\tau_1, \tau_2 \to \infty} P_{2e} = \epsilon$ as expected.

With $\tau_1 = \tau_2 = \tau/2$, we obtain the average net work as
\begin{equation}
\langle W_\mathrm{net} \rangle = \frac{(q-\epsilon)(1-e^{-\tau/2})^2}{1-e^{-\tau}}
\left[ X(T_1,q) - X(T_2,\epsilon) \right] \,,
\label{eq:Wnet_for_tau}
\end{equation}
so the monotonically increasing factor $(1-e^{-\tau/2})^2/(1-e^{-\tau})$ for the time scale $\tau$ is decoupled from the rest of the formula and only plays the role of an overall factor. It is important to note that the decoupling holds regardless of the  $\tau_1 = \tau_2$ condition; the overall factor becomes $(1-e^{-\tau_1})(1-e^{-\tau_2})/(1-e^{-\tau_1+\tau_2})$. 
The average power output $\langle P \rangle$ is given by
\begin{equation}
\langle P \rangle = \frac{(q-\epsilon)(1-e^{-\tau/2})^2}{\tau(1-e^{-\tau})}
\left[ X(T_1,q) - X(T_2,\epsilon) \right] \,,
\label{eq:power}
\end{equation}
which decreases monotonically with $\tau$.
Therefore, from now on, we discard the time dependence altogether and focus on other parameters,
i.e., denoting
\begin{equation}
\langle W_\mathrm{net} \rangle \equiv \langle P \rangle \equiv (q-\epsilon) \left[ X(T_1,q) - X(T_2,\epsilon) \right] \,,
\label{eq:W_P_equally}
\end{equation}
without considering the overall factor involving $\tau$
for notational convenience.
Numerically, we obtain the net work and efficiency for $(q, \epsilon)$ combination, as shown in Fig.~\ref{fig:Wnet_and_eta_3D}.
In Sec.~\ref{sec:optimal_power}, we derive the condition for the efficiency at the maximum power output.

\subsection{Efficiency at the maximum power output}
\label{sec:optimal_power}

\subsubsection{The condition for the maximum power output}
\label{sec:exact_formalism}

For a given $T_2/T_1$ value, the maximum power output condition for the two-variable function is
\begin{equation}
\displaystyle \left. \frac{\partial \langle P \rangle}{\partial q} \right|_{q=q^*,\epsilon=\epsilon^*} = \left. \frac{\partial \langle P \rangle}{\partial \epsilon} \right|_{q=q^*,\epsilon=\epsilon^*} = 0  \,,
\label{eq:first_derivative_condition}
\end{equation}
which leads to
\begin{subequations}
\begin{equation}
\displaystyle 1 - \frac{X(T_2,\epsilon^*)}{X(T_1,q^*)} = \frac{q^* - \epsilon^*}{q^* (1-q^*) \ln[(1-q^*)/q^*]} \,,
\label{eq:optimal_condition_for_q}
\end{equation}
and
\begin{equation}
\displaystyle 1 - \frac{X(T_2,\epsilon^*)}{X(T_1,q^*)} = \frac{(T_2 / T_1)(q^* - \epsilon^*)}{\epsilon^* (1-\epsilon^*) \ln[(1-q^*)/q^*]} \,,
\label{eq:optimal_condition_for_epsilon}
\end{equation}
\end{subequations}
from Eq.~\eqref{eq:W_P_equally}.
By eliminating the left-hand side of Eqs.~\eqref{eq:optimal_condition_for_q} and \eqref{eq:optimal_condition_for_epsilon}, we obtain
the following simple relation
\begin{subequations}
\begin{equation}
\frac{T_2 q^* (1-q^*)}{T_1 \epsilon^* (1-\epsilon^*)} = 1 \,,
\label{eq:global_optimum_condition}
\end{equation}
or
\begin{equation}
\epsilon^* = \frac{1}{2} \left[ 1 -  U(\eta_C,q^*) \right] \,,
\label{eq:global_optimum_condition_solution}
\end{equation}
\end{subequations}
with
\begin{equation}
U(\eta_C,q^*) \equiv \sqrt{4\eta_C q^* (1-q^*) + (1 - 2q^*)^2} \,.
\label{eq:U_definition}
\end{equation}
By substituting $\epsilon^*$ as a function of $q^*$ in Eq.~\eqref{eq:global_optimum_condition_solution} to Eq.~\eqref{eq:optimal_condition_for_q} or Eq.~\eqref{eq:optimal_condition_for_epsilon},
we get the optimum condition
\begin{equation}
\begin{aligned}
 & \ln \left( \frac{\displaystyle 1-q^*}{\displaystyle q^*} \right) - \frac{\displaystyle T_2}{\displaystyle T_1} \ln \left[ \frac{\displaystyle 1 + U(\eta_C,q^*)}{\displaystyle 1 - U(\eta_C,q^*)} \right] \\
 & - \frac{\displaystyle q^* - \frac{1}{2} + \frac{1}{2}U(\eta_C,q^*)}{\displaystyle q^*(1-q^*)} = 0 \,.
\end{aligned}
\label{eq:f_q_star}
\end{equation}
Furthermore, the condition in Eq.~\eqref{eq:f_q_star} leads to the following form of $\eta_\mathrm{op}$ from Eq.~\eqref{eq:eta},
\begin{equation}
\eta_\mathrm{op} = \frac{\displaystyle q^* - \frac{1}{2} + \frac{1}{2} U(\eta_C,q^*)}{\displaystyle q^* (1-q^*) \ln [(1-q^*)/q^*]} \,.
\label{eq:optimal_eta_for_q_star_T}
\end{equation}
It is also straightforward to show that this point $(q^*,\epsilon^*)$ is indeed a maximum point by investigating the second derivatives of the power.

In order to calculate the efficiency for given $T_2/T_1$ at the maximum power, first find the $q^*$ value satisfying Eq.~\eqref{eq:f_q_star} and substitute the $q^*$ value to Eq.~\eqref{eq:optimal_eta_for_q_star_T}.
As Eq.~\eqref{eq:f_q_star} is a transcendental equation, the closed-form solution for $\eta_\mathrm{op}$ is unattainable.

\subsubsection{Asymptotic behaviors obtained from series expansion}
\label{sec:series_expansion}

\begin{figure}
\includegraphics[width=0.45\textwidth]{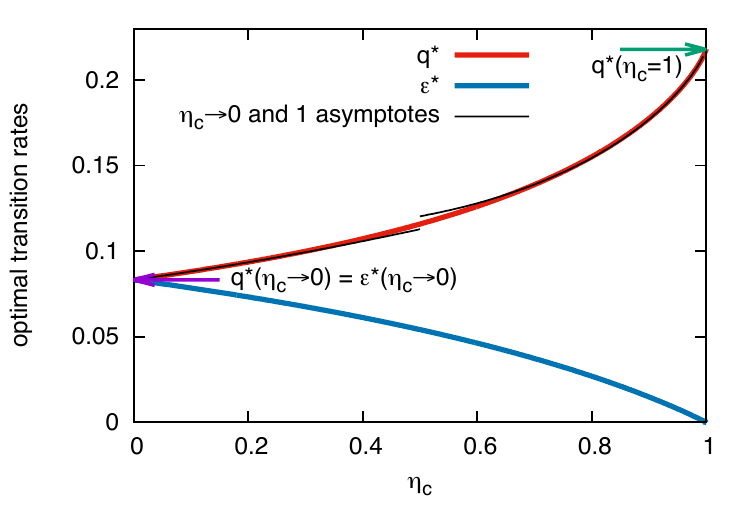}
\caption{Numerically found $q^*$ and $\epsilon^*$ values satisfying Eq.~\eqref{eq:f_q_star}, as a function of $\eta_C = 1 - T_2/T_1$,
along with the $q^* (\eta_C \to 0) = \epsilon^* (\eta_C \to 0)$ and $q^* (\eta_C = 1)$ values presented in Sec.~\ref{sec:series_expansion}. $\epsilon^* (\eta_C = 1) = 0$ (the horizontal axis). The $\eta_C \to 0$ asymptote indicates Eq.~\eqref{eq:q_series_expansion_wrt_eta_C} up to the $\eta_C^2$ term, 
and the $\eta_C \to 1$ asymptote indicates Eq.~\eqref{eq:q_series_expansion_wrt_eta_C_near_one} up to the $(1-\eta_C)$ term with the coefficients given by Eqs.~\eqref{eq:b_ln_solution} and \eqref{eq:b_1_solution}.}
\label{fig:optimal_qstar_estar}
\end{figure}

\begin{figure}
\includegraphics[width=0.9\columnwidth]{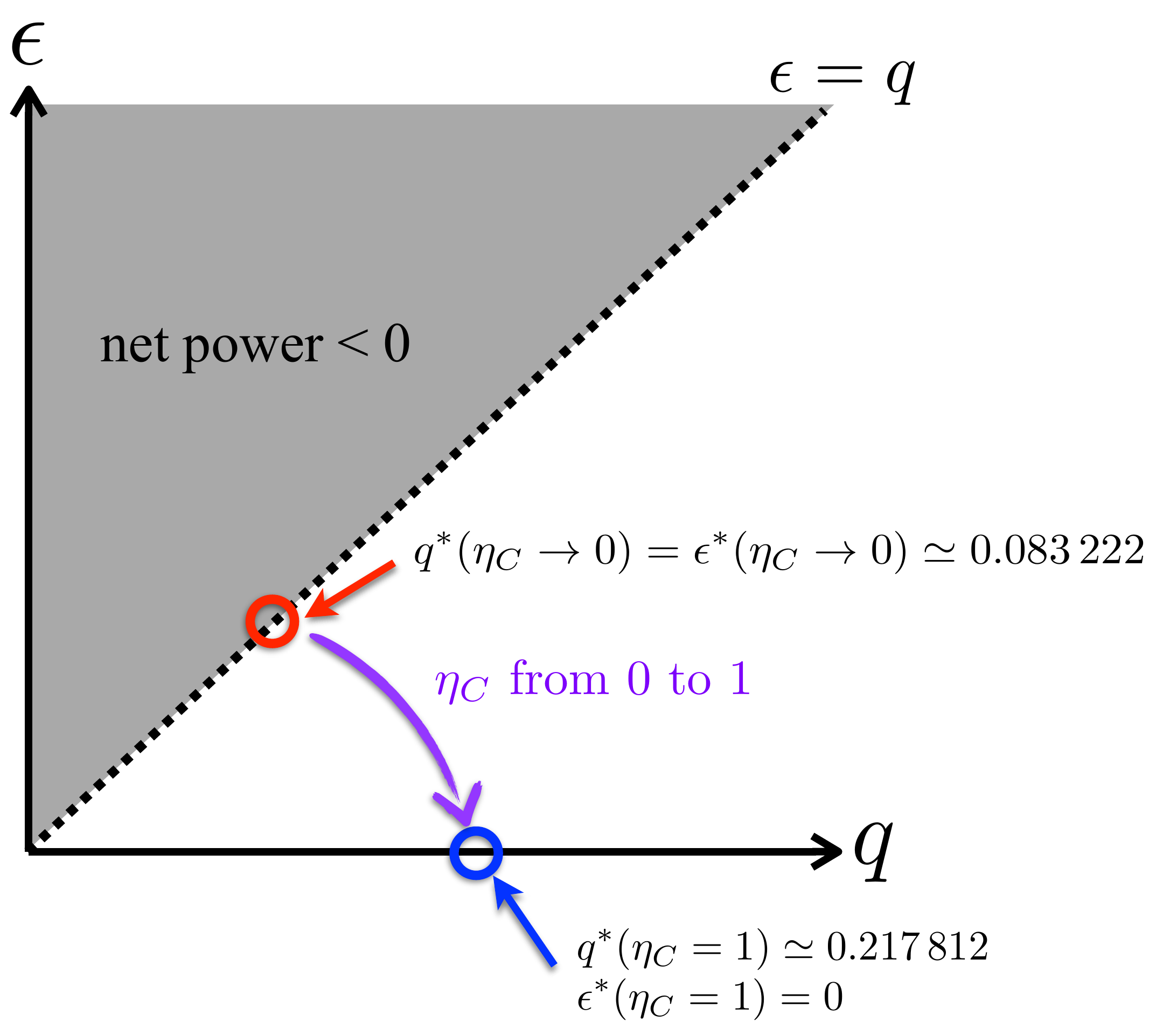}
\caption{Illustration of the optimal transition rates $(q^*,\epsilon^*)$ for the maximum power output as the $T_2/T_1$ value varies.} 
\label{fig:optimal_q_e_illustration}
\end{figure}

The upper bound for $q^*$ is given by the condition $\eta_C = 1$, satisfying $\ln[(1-q^*)/q^*] = 1/(1-q^*)$ and $q^* (\eta_C = 1) \simeq 0.217\,812$ found numerically and $\epsilon^* (\eta_C = 1) = 0$ exactly from Eq.~\eqref{eq:global_optimum_condition_solution}. $\eta_C = 0$ always satisfies Eq.~\eqref{eq:f_q_star} regardless of $q^*$ values,
so finding the optimal $q^*$ is meaningless (in fact, when $\eta_C = 0$, the operating regime for the engine is shrunk to the line $q = \epsilon$ and there cannot be any positive work).
Therefore, let us examine the case $\eta_C \simeq 0$ using the series expansion of $q^*$ with respect to $\eta_C$, as
\begin{equation}
q^* = a_0 + a_1 \eta_C + a_2 \eta_C^2 + a_3 \eta_C^3 + \mathcal{O}\left({\eta_C^4}\right) \,.
\label{eq:q_series_expansion_wrt_eta_C}
\end{equation}
Substituting Eq.~\eqref{eq:q_series_expansion_wrt_eta_C} into Eq.~\eqref{eq:f_q_star} and expanding the left-hand side with respect to $\eta_C$ again, we obtain
\begin{equation}
c_1 \eta_C + c_2 \eta_C^2 + c_3 \eta_C^3 + \mathcal{O}\left({\eta_C^4}\right) = 0 \,,
\label{eq:f_star_expansition_wrt_eta_C}
\end{equation}
where $c_n$ describes the relation among $a_0, \cdots, a_{n-1}$, each of which should be identically zero to satisfy Eq.~\eqref{eq:f_star_expansition_wrt_eta_C}. 
Letting the linear coefficient $c_1$ be zero yields
\begin{equation}
\frac{2}{1-2a_0} = \ln \left( \frac{1-a_0}{a_0} \right) \,,
\label{eq:a_0_expression}
\end{equation}
from which the lower bound for $q^* (\eta_C \to 0) = a_0 = \epsilon^* (\eta_C \to 0) \simeq 0.083\,222$ found numerically [$\lim_{\eta_C \to 0} U(\eta_C,q^*) = 1 - 2q^*$, thus $\epsilon^* (\eta_C \to 0) = q^* (\eta_C \to 0)$ by Eq.~\eqref{eq:global_optimum_condition_solution}].
Figure~\ref{fig:optimal_qstar_estar} shows the numerical solution $(q, \epsilon) = (q^*, \epsilon^*)$ as a function of $\eta_C$,
where the asymptotic behaviors derived above hold when $\eta_C \simeq 0$ and $\eta_C \simeq 1$.
It seems that $q^*$ is monotonically increased and $\epsilon^*$ is monotonically decreased, as $\eta_C$ is increased,
i.e., $q_\mathrm{min}^* = q^* (\eta_C \to 0)$, $q_\mathrm{max}^* = q^* (\eta_C = 1)$,
$\epsilon_\mathrm{min}^* = 0$, and $\epsilon_\mathrm{max}^* = \epsilon^* (\eta_C \to 0)$.
Figure~\ref{fig:optimal_q_e_illustration} illustrates the situation on the $(q,\epsilon)$ plane.
The linear coefficient $a_1$ in Eq.~\eqref{eq:q_series_expansion_wrt_eta_C} can be written in terms of $a_0$ when we let $c_2 = 0$ in Eq.~\eqref{eq:f_star_expansition_wrt_eta_C}, 
and the quadratic coefficient $a_2$ in Eq.~\eqref{eq:q_series_expansion_wrt_eta_C} can also be written in terms of $a_0$ alone, by letting $c_3 = 0$ in Eq.~\eqref{eq:f_star_expansition_wrt_eta_C} and using the relations in Eqs.~\eqref{eq:a_0_expression} and $a_1$ expressed by $a_0$ terms, 
which are well consistent with the numerical solution as shown in Fig.~\ref{fig:optimal_qstar_estar}.

\begin{figure}
\includegraphics[width=\columnwidth]{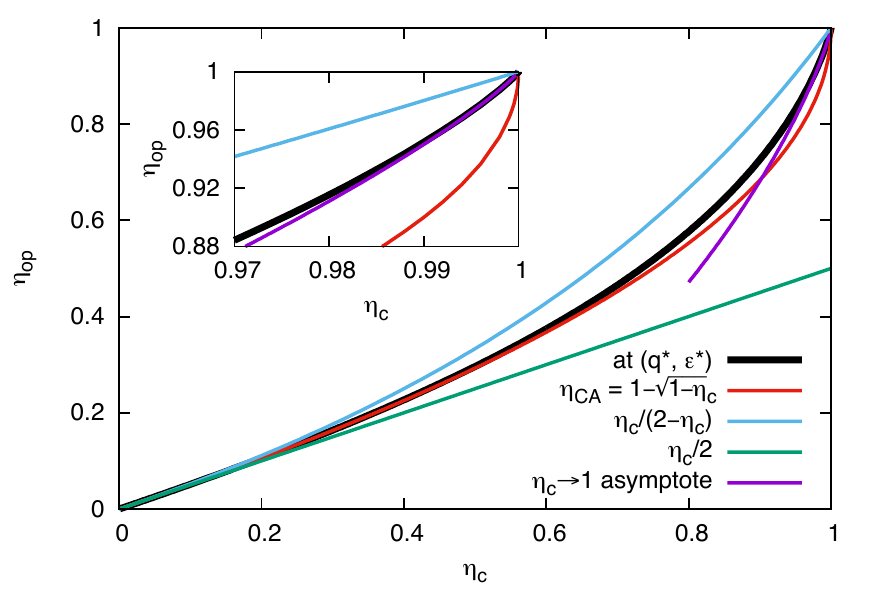}
\caption{The efficiency at the maximum power $\eta_\mathrm{op}$ as the function of the Carnot efficiency $\eta_C$ in Eq.~\eqref{eq:optimal_eta_for_q_star_T} using numerically found optimal $q^*$ values, along with various asymptotic cases: the Curzon-Ahlborn efficiency $\eta_\mathrm{CA}$ in Eq.~\eqref{eq:eta_CA}, 
the upper bound $\eta_C / (2-\eta_C)$ and the lower bound $\eta_C/2$ in Ref.~\cite{Esposito2010},
and the $\eta_C \to 1$ asymptote 
for $\eta_C \ge 0.8$.
The inset shows the region $0.97 < \eta_C < 1$.
}
\label{fig:optimal_eta}
\end{figure}

With the relations of coefficients in hand, we find the asymptotic behavior of $\eta_\mathrm{op}$ in Eq.~\eqref{eq:optimal_eta_for_q_star_T} by expanding it with respect to $\eta_C$ after substituting $q^*$ as the series expansion of $\eta_C$ in Eq.~\eqref{eq:q_series_expansion_wrt_eta_C}.
Then,
\begin{equation}
\eta_\mathrm{op} = \frac{1}{2} \eta_C + \frac{1}{8} \eta_C^2 + \frac{7-24a_0+24a_0^2}{96(1-2a_0)^2} \eta_C^3 + \mathcal{O}\left({\eta_C^4}\right) \,.
\label{eq:eta_op_expansion_wrt_eta_C_as_CA}
\end{equation}
With this method, we are able to find the coefficients in terms of $a_0$ up to an arbitrary order in principle.
We would like to emphasize that the expansion form of $\eta_\mathrm{op}$ in Eq.~\eqref{eq:eta_op_expansion_wrt_eta_C_as_CA} has exactly the same coefficients up to the quadratic term to those of the Curzon-Ahlborn efficiency~\cite{Chambadal1957,Novikov1958,Curzon1975} defined as
\begin{equation}
\eta_\mathrm{CA} = 1 - \sqrt{T_2/T_1} = 1 - \sqrt{1 - \eta_C} \,,
\label{eq:eta_CA}
\end{equation}
with the expansion form
\begin{equation}
\eta_\mathrm{CA} = \frac{1}{2} \eta_C + \frac{1}{8} \eta_C^2 + \frac{1}{16} \eta_C^3 + \frac{5}{128} \eta_C^4 + \mathcal{O}(\eta_C^5) \,,
\label{eq:eta_CA_expansion}
\end{equation}
when $\eta_C \simeq 0$.
As a result, numerically found $\eta_\mathrm{op}$ by solving Eq.~\eqref{eq:f_q_star} and substituting the $q^*$ value to Eq.~\eqref{eq:optimal_eta_for_q_star_T}, and $\eta_\mathrm{CA}$ share a very similar functional form for $\eta_C \lesssim 1/2$, as shown in Fig.~\ref{fig:optimal_eta}. In fact, the linear term $\eta_C / 2$ and quadratic term $\eta_C^2 / 8$ are naturally from the strong coupling between the thermodynamic fluxes and the symmetry between the reservoirs (as we will check in Sec.~\ref{sec:entropy_production}, the reservoir symmetry is related to the symmetry in the entropy production at the hot and cold reservoir and holds only approximately in our model)~\cite{VanDenBroeck2005,Esposito2009}.
The third order coefficient ($\simeq 0.077\,492$) in Eq.~\eqref{eq:eta_op_expansion_wrt_eta_C_as_CA}, however, is
clearly different from $1/16$ for the $\eta_\mathrm{CA}$. 
In other words, the deviation from $\eta_\mathrm{CA}$ for $\eta_\mathrm{op}$ enters from the third order that has not been theoretically investigated yet.
Indeed, $\eta_\mathrm{op}$ deviates from $\eta_\mathrm{CA}$ for $\eta_C \gtrsim 1/2$, until they coincide at $\eta_C = 1$.
Therefore, the efficiency $\eta_\mathrm{op}$ of our model at maximum power output is different from $\eta_\mathrm{CA}$.

For $\eta_C \simeq 1$, we need to consider the logarithmic correction due to the functional form, based on the numerical evidence also shown in Fig.~\ref{fig:optimal_eta}. In contrast to the linear heat conduction for the Curzon-Ahlborn endoreversible engine~\cite{Chambadal1957,Novikov1958,Curzon1975}, our model has an exponential or Boltzmann type of relaxation process. We believe that this different functional form of heat conduction process results in the different types of singularity at $\eta_C \simeq 1$: the algebraic singularity of $\eta_\mathrm{CA}$ in Eq.~\eqref{eq:eta_CA} at $\eta_C = 1$ with the infinite slope, and the logarithmic singularity in our case. We take the singular series expansion of the functional form in Eq.~\eqref{eq:f_q_star} near $\eta_C = 1$ as
\begin{equation}
\begin{aligned}
q^* = & q_\mathrm{max}^* + b_{\ln} (1-\eta_C) \ln(1-\eta_C) \\
 & + b_1 (1-\eta_C) + \mathcal{O}\left[{(1-\eta_C)^2}\right] \,.
\end{aligned}
\label{eq:q_series_expansion_wrt_eta_C_near_one}
\end{equation}
It is possible to consider other types of terms such as $(1-\eta_C) \ln^2 (1-\eta_C)$, but we will check that it is enough to
predict the functional form of $\eta_\mathrm{op}$, consistent with an alternative approach
from entropy-production-based analysis provided in Sec.~\ref{sec:entropy_production}.
If we take only the zeroth order term, we obtain the identity
\begin{equation}
\frac{1}{1-q_\mathrm{max}^*} = \ln\left( \frac{1-q_\mathrm{max}^*}{q_\mathrm{max}^*} \right) \,,
\label{eq:qmax_solution}
\end{equation}
exactly at $\eta_C = 1$ that is already mentioned in the first part of this subsection.
Similar to the $\eta_C \simeq 0$ case, by letting each coefficient be zero, we find the relations among the coefficients as
\begin{equation}
b_{\ln} = q_\mathrm{max}^* (1-q_\mathrm{max}^*)^2 \,,
\label{eq:b_ln_solution}
\end{equation}
and
\begin{equation}
b_1 = q_\mathrm{max}^* (1-q_\mathrm{max}^*)^2 \left\{ 1 + \ln[q_\mathrm{max}^*(1-q_\mathrm{max}^*)] \right\} \,,
\label{eq:b_1_solution}
\end{equation}
which are well consistent with the numerical solution as shown in Fig.~\ref{fig:optimal_qstar_estar}.

Again, the asymptotic behavior of $\eta_\mathrm{op}$ in Eq.~\eqref{eq:optimal_eta_for_q_star_T} for $\eta_C \simeq 1$ can be deduced from the series expansion in terms of $(1-\eta_C) > 0$, using Eqs.~\eqref{eq:q_series_expansion_wrt_eta_C_near_one}, \eqref{eq:b_ln_solution}, and \eqref{eq:b_1_solution}, which is
\begin{equation}
\begin{aligned}
\eta_\mathrm{op} = & 1 + (1-q_\mathrm{max}^*) (1 - \eta_C) \ln (1 - \eta_C) \\
 & + (1-q_\mathrm{max}^*) \ln[q_\mathrm{max}^*(1-q_\mathrm{max}^*)] (1 - \eta_C) \\
 & + \mathcal{O}\left[ (1-\eta_C )^2\right] \,,
\end{aligned}
\label{eq:eta_op_expansion_near_eta_C_1}
\end{equation}
based on the relations in Eqs.~\eqref{eq:b_ln_solution} and \eqref{eq:b_1_solution}
[the same procedure as the one leading to \eqref{eq:eta_op_expansion_wrt_eta_C_as_CA}].
As shown in Fig.~\ref{fig:optimal_eta}, however, the asymptotic form 
only holds in a rather limited range of $\eta_C$ very close to unity, indicating the necessity to taking higher order terms into account for more accurate asymptotic behavior.

\subsection{The entropy production relation}
\label{sec:entropy_production}

\begin{figure}
\includegraphics[width=0.9\columnwidth]{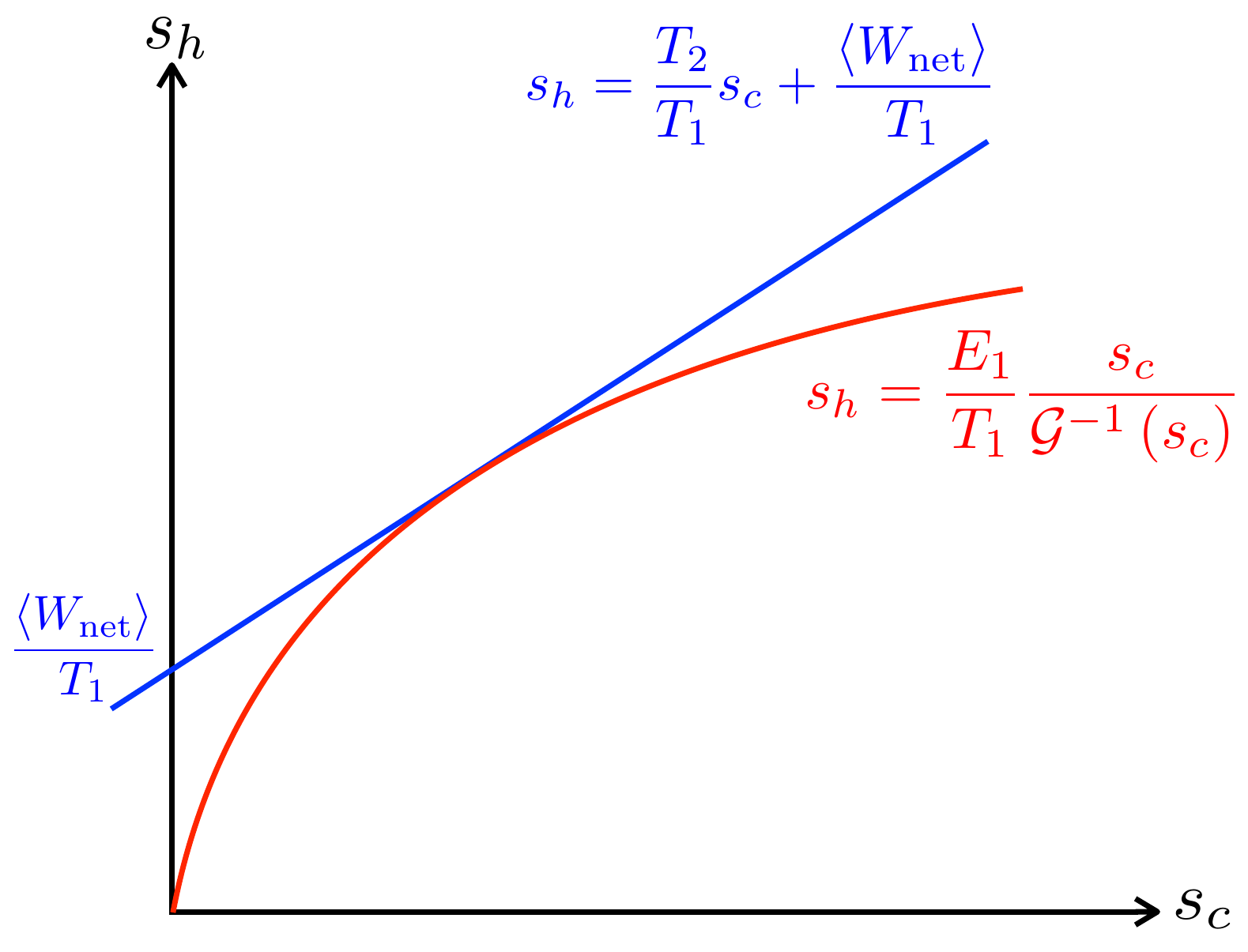}
\caption{The entropy relation between $s_h$ and $s_c$, given by Eq.~\eqref{eq:entropy_relation_our_model} and the linear relation in Eq.~\eqref{eq:entropy_first_law} representing the first law of thermodynamics. The maximum value of $\langle W_\mathrm{net} \rangle$ (the intercept of the linear relation on the vertical axis times $T_1$) is achieved when the line becomes the tangential one of the curve, as illustrated here.}
\label{fig:entropy_relation}
\end{figure}

In Ref.~\cite{JMPark2016}, it is argued that the necessary and sufficient condition for the Curzon-Ahlborn efficiency at the maximum power is that the entropy productions at the hot and cold reservoirs (denoted by $s_h$ and $s_c$, respectively, ) should be related by a specific functional form, namely, $s_h = \mathcal{F}(s_c)$ where $\mathcal{F}(x) = x / (1+\zeta x)$ with the system-specific constant $\zeta$.

The entropy production in our model is given by
\begin{equation}
\begin{aligned}
s_h = \frac{\langle Q_1 \rangle}{T_1} \equiv (q-\epsilon) \frac{E_1}{T_1} \,, \\
s_c = \frac{\langle Q_2 \rangle}{T_2} \equiv (q-\epsilon) \frac{E_2}{T_2} \,,
\end{aligned}
\label{eq:entropy_production_form_for_our_model}
\end{equation}
where we again discard the common explicit time dependent term $(1-e^{-\tau/2})^2/(1-e^{-\tau})$, which does not affect the following discussion for notational convenience.
Given $T_1$ and $T_2$ and putting $E_1$ as a constant, we obtain the entropy relation given by
\begin{equation}
s_h = \frac{E_1}{T_1}\frac{s_c}{\mathcal{G}^{-1}\left(s_c\right)} \,,
\label{eq:entropy_relation_our_model}
\end{equation}
where $\mathcal{G}^{-1}$ is the inverse function of $\mathcal{G}$ defined
from the relation in Eq.~\eqref{eq:entropy_production_form_for_our_model}, $s_c = \mathcal{G}(E_2/T_2)$.
Note that $s_h$ is an increasing function with respect to $s_c$
while $d s_h/d s_c$ is a decreasing one,
so $s_h (s_c)$ is an increasing and concave function of $s_c$ as illustrated in Fig.~\ref{fig:entropy_relation}.
Therefore, the unique [guaranteed by the concavity of $s_h (s_c)$] optimal entropy production denoted by $s_c^*$,
which makes the power reach its maximum value, is determined by
\begin{equation}
\left. \frac{d s_h}{d s_c}\right|_{s_c = s_c^*}
= \frac{E_1}{T_1} \frac{\epsilon^* \tilde{\epsilon}^*}{q -\epsilon
+ \mathcal{G}^{-1}\left(s_c^* \right) \epsilon^* \tilde{\epsilon}^*}
=\frac{T_2}{T_1},
\label{eq:noh}
\end{equation}
where $T_2/T_1$ comes from the thermodynamic first law,
\begin{equation}
s_h = \frac{T_2}{T_1} s_c + \frac{\langle W_{\rm net} \rangle}{T_1} \,.
\label{eq:entropy_first_law}
\end{equation}
The parameter $\epsilon^*$ obtained from Eq.~\eqref{eq:noh} is still a function of $E_1$ or $q$.
By optimizing the entropy production with respect to $q$, we find the same optimal 
$q^*$ and $\epsilon^*$  as those in the previous section.

As $\mathcal{G}^{-1}(s_c)$ is not a linear function of $s_c$, we do not have the relation $s_h = \mathcal{F}(s_c)$ mentioned before, so the fact that
$\eta_\mathrm{op} \neq \eta_\mathrm{CA}$ is consistent with Ref.~\cite{JMPark2016}.
However, we reveal that it is possible to find the regime where the entropy production of our model approximately follows the functional form $\mathcal{F}(x)$,
which indeed corresponds to the $\eta_\mathrm{op} \simeq \eta_\mathrm{CA}$ regime,
as we show in the following section.

\subsubsection{The linear regime}
\label{sec:entropy_production_linear_regime}

First, we take the regime where $\Delta T= T_1-T_2 \ll T_1$.
Then from Eq.~\eqref{eq:noh}, one can see the solution $\Delta E^* = E_1 - E_2^* \ll E_1$ or $s_c^* \ll 1$,
which allows the small $s_c$ expansion of $\mathcal{G}^{-1}$ up to the linear order as
\begin{equation}
\mathcal{G}^{-1}(s_c) \simeq \mathcal{G}^{-1}(0)
+ \left. \frac{d \mathcal{G}^{-1}}{d s_c} \right|_{s_c = 0} s_c = \frac{E_1}{T_1} + \zeta(E_1/T_1) s_c \,,
\label{eq:expansion}
\end{equation}
where the constant $\zeta(E_1/T_1)$ is given by
\begin{equation}
\zeta = 2 \left(\frac{T_1}{E_1}\right)^2 \left[1 + \cosh\left(\frac{E_1}{T_1}\right)\right] \,.
\label{eq:zeta_form}
\end{equation}
Inserting Eq.~\eqref{eq:expansion} to the entropy relation in Eq.~\eqref{eq:entropy_relation_our_model},
we find that the entropy production for hot and cold reservoirs actually follows the functional form $\mathcal{F}(x) = x / (1+\zeta x)$ for $\eta_C \to 0$,
which explains the Curzon-Ahlborn-like behavior for $\eta_C \to 0$.
We have already checked that $\eta_\mathrm{op} \simeq \eta_\mathrm{CA}$ as $\eta_C \to 0$ due to the same linear and quadratic coefficients from the series expansion in Sec.~\ref{sec:series_expansion}, but the entropy production relation suggests that there could be a deeper relation between our model and engines with the optimal efficiency of $\eta_\mathrm{CA}$ than the reasons for the linear and quadratic coefficients, namely, the strong coupling between the thermodynamics fluxes and the reservoir symmetry, respectively. In fact, the implication of the reservoir symmetry in the expression
$\mathcal{F}^{-1} (x) = \mathcal{F}(-x)$ holds only for $\eta_C \to 0$.

We emphasize that the behavior of efficiency $\eta_\mathrm{op} \simeq \eta_\mathrm{CA}$ in $\eta_C \to 0$
is independent of the value $q$.
However we can optimize $\langle W_{\rm net} \rangle$ with respect to $q$ as following.
The optimal condition to maximize $\langle W_\mathrm{net} \rangle$
for a given $T_2/T_1$ value is equivalent to minimize $\zeta$, because
\begin{equation}
\langle W_\mathrm{net} \rangle = \frac{T_1}{\zeta} \left(1 - \sqrt{\frac{T_2}{T_1}}\right)^2 \propto 1/\zeta  \,,
\label{eq:Wnet_zeta_relation}
\end{equation}
from the condition of tangent $T_2 / T_1 = 1/(1+\zeta s_c^*)^2$.
The condition for the minimum value of $\zeta$ for given $E_1/T_1$ is, by taking the derivative of the functional form in Eq.~\eqref{eq:zeta_form} with respect to $E_1/T_1$ and using the relation $q/(1-q) = e^{-E_1/T_1}$,
\begin{equation}
\frac{2}{1-2q} =\ln\left(\frac{1-q}{q}\right) \,,
\label{eq:zeta_minimum_condition}
\end{equation}
which is equivalent to the condition for the zeroth order term of $q^* (\eta_C \to 0)$ given by Eq.~\eqref{eq:a_0_expression}.
In other words, the optimal condition, at least for the lowest order of $\eta_C$ at $\eta_C \simeq 0$, can also be derived from
the functional form of entropy production given by Eqs.~\eqref{eq:entropy_relation_our_model} and \eqref{eq:zeta_form},
further supporting the consistency of our result.

\subsubsection{The logarithmic regime}
\label{sec:entropy_production_log_regime}

The other extreme regime is $T_2 \ll T_1$, or the $\eta_C \to 1$ limit, where
the solution satisfying Eq.~\eqref{eq:noh} exists in the region of $E_2^*/T_2 \gg 1 $ or $s_c \gg 1$.
In this limit we rewrite the relation between $s_c$ and $E_2/T_2$ in
Eq.~\eqref{eq:entropy_production_form_for_our_model} as
\begin{equation}
\frac{E_2}{T_2} \simeq \frac{s_c}{q} \left( 1+  \frac{1}{q} e^{-E_2 / T_2} \right) \,.
\end{equation}
Using the above, we obtain $\mathcal{G}^{-1}(s_c)$ up to the order of $(s_c/q^2) e^{-s_c/q}$
\begin{equation}
\mathcal{G}^{-1}(s_c) \simeq \frac{s_c}{q} + \frac{s_c}{q^2} e^{-s_c / q} \,.
\label{eq:exp_correction}
\end{equation}
Inserting Eq.~\eqref{eq:exp_correction} to Eq.~\eqref{eq:entropy_relation_our_model}, the entropy production
relation reads
\begin{equation}
s_h = \frac{E_1}{T_1} \frac{q^2}{q + e^{-s_c / q} } \,,
\label{eq:entropy_exp}
\end{equation}
and using the condition 
\begin{equation}
\left. \frac{d s_h}{d s_c} \right|_{s_c = s_c^*} = \frac{E_1}{T_1} \frac{qe^{-s_c/q}}{(q + e^{-s_c/q})^2} \simeq \frac{E_1 e^{-s_c/q}}{q T_1} = \frac{T_2}{T_1} \,, 
\label{eq:slope_condition_for_log_limit}
\end{equation}
we derive the efficiency at the maximum power,
\begin{equation}
\begin{aligned}
\eta_{\mathrm{op}} \simeq 1 - & \frac{T_1}{E_1}\left(1-\eta_C \right)
\left[ 1+ \frac{T_1}{E_1} \left(1-\eta_C \right) \right] \\
& \times \ln \left[ \frac{E_1}{qT_1 \left(1-\eta_C \right)} \right] \,.
\end{aligned}
\label{eq:efficiency_log}
\end{equation}

Optimizing the power with respect to $q$ or $E_1$ is equivalent to the maximizing the saturation value
of Eq.~\eqref{eq:entropy_exp}, which is $\lim_{s_c \to \infty} s_h = q E_1/T_1$.
This condition yields the functional form $q^*_{\rm max}$ should satisfy, which is equivalent to Eq.~\eqref{eq:qmax_solution}.
Using the result $E_1/T_1 = (1-q^{*}_{\rm max})^{-1}$,
we obtain $\eta_{\rm op}$ in the $(q,\epsilon)$ space given by
\begin{equation}
\begin{aligned}
\eta_{\mathrm{op}} \simeq 1 + & \left( 1-q^{*}_{\rm max}\right) \left(1-\eta_C \right)
\left[ 1+ \left( 1-q^{*}_{\rm max} \right)  \left(1-\eta_C \right) \right] \\
 & \times\ln \left[ q^{*}_{\rm max} \left(1-q^{*}_{\rm max}\right) \left(1-\eta_C \right) \right],
\end{aligned}
\label{eq:eta_op_near_eta_C_1_from_entropy}
\end{equation}
which is consistent with the result in the previous section; in fact, Eq.~\eqref{eq:eta_op_near_eta_C_1_from_entropy} includes higher order terms,
namely, $(1 - \eta_C)^2$ and $(1-\eta_C)^2 \ln (1-\eta_C)$,
than Eq.~\eqref{eq:eta_op_expansion_near_eta_C_1}.
We also emphasize that this consistency justifies the series expansion form in Eq.~\eqref{eq:q_series_expansion_wrt_eta_C_near_one}.

Based on the analysis above for the regime where $T_2 \ll T_1$, we reach the conclusion that there cannot be a $q$ value making
the higher order terms than the linear term in the expansion of $\mathcal{G}^{-1}$ for any given $T_1$ and $T_2$ values. In other words, one again
can see the systematic difference between our model and the models belonging to the Curzon-Ahlborn efficiency at the maximum power.

\section{Extension to multi-level heat engine}
\label{sec:three_level}

\begin{figure}
\includegraphics[width=0.5\columnwidth]{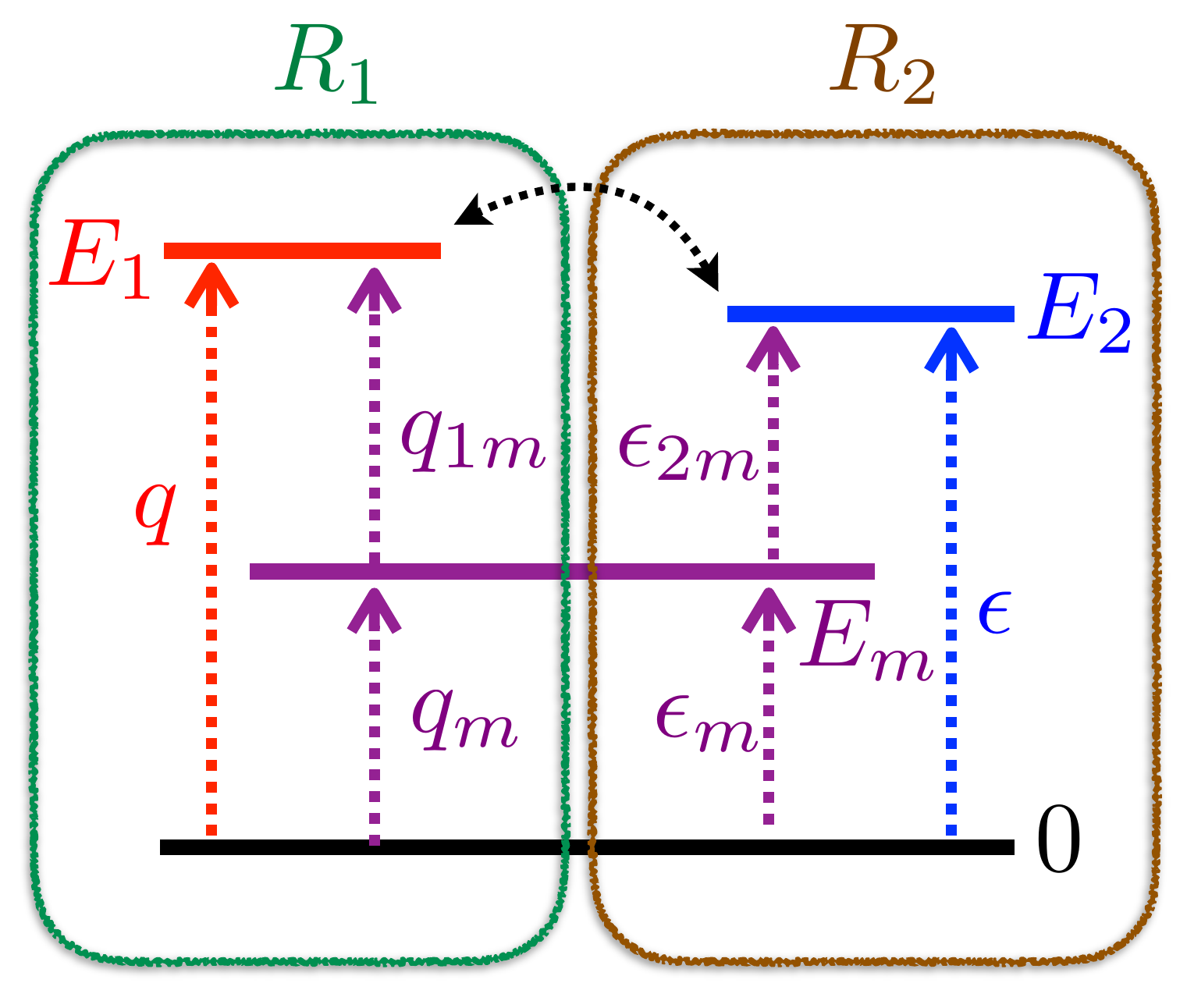}
\caption{The energy levels for each reservoir in the three-level heat engine, along with the transition rates, are illustrated. The work is extracted by adjusting the most upper level (the second excited state) between the reservoir contacts.}
\label{fig:three_level_heat_engine}
\end{figure}

\begin{figure*}
\begin{tabular}{lll}
(a) & (b) & \\
\includegraphics[height=0.3\textwidth]{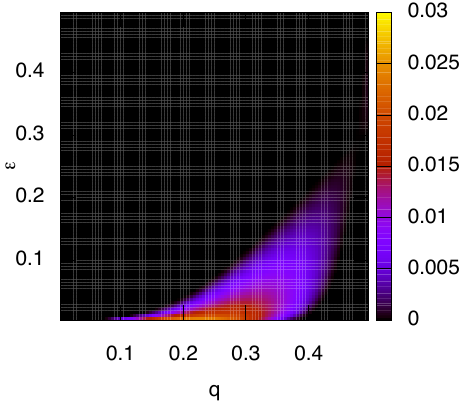} &
\includegraphics[height=0.3\textwidth]{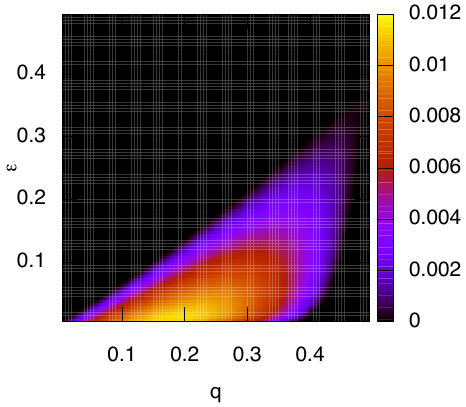} \\
\end{tabular}
\caption{The average power output $\langle W_\mathrm{net} \rangle / \tau$ for the three-level system with the parameters $T_1 = 1$, $T_2 = 0.1$, $E_m = 0.5$, and (a) $\tau = 1$ [the maximum value of $\langle W_\mathrm{net} \rangle / \tau$ occurs at $q^* = 0.215(5)$ and $\epsilon^* = 0.005(5)$], and (b) $\tau = 10.0$ [$q^* = 0.185(5)$ and $\epsilon^* = 0.010(5)$]. For better visibility focused on the $\langle W_\mathrm{net} \rangle / \tau \ge 0$ regime, we set all of the negative values as $0$.
We take the normalization convention $q + \tilde{q} = q_m + \tilde{q}_m = q_{1m} + \tilde{q}_{1m} = \epsilon + \tilde{\epsilon} = \epsilon_m + \tilde{\epsilon} = \epsilon_{2m} + \tilde{\epsilon}_{2m} = 1$.}
\label{fig:three_level_example}
\end{figure*}

Finally, we would like to remark on the possible extension to multi-level systems, i.e., systems with more than two levels, which are more general cases. In that framework, our two-level heat engine can be taken as a simplified one considering only the ground and first excited states. For simplicity, again we assume two heat reservoirs $R_1$ and $R_2$, which are characterized by the temperatures $T_1$ and $T_2$, and contacted with the system during the time $\tau_1$ and $\tau_2$, respectively. 

First, let us take the three-level system, where we consider the ground, first excited, and second excited states for each reservoir. We further simplify the situation by differentiating only the second excited states of the reservoirs, namely $E_1$ for $R_1$ and $E_2$ for $R_2$, and the common value $E_m$ for the first excited state, as depicted in Fig.~\ref{fig:three_level_heat_engine}. The transition rates are denoted by $q$ (the ground state to $E_1$ in $R_1$), $q_m$ (the ground state to $E_m$ in $R_1$), $q_{1m}$ ($E_m$ to $E_1$ in $R_1$), $\epsilon$ (the ground state to $E_2$ in $R_2$), $\epsilon_m$ (the ground state to $E_m$ in $R_2$), and $\epsilon_{2m}$ ($E_m$ to $E_2$ in $R_2$); their reverse transition rates are $\tilde{q}$, $\tilde{q}_{m}$, $\tilde{q}_{1m}$, $\tilde{\epsilon}$, $\tilde{\epsilon}_m$, and $\tilde{\epsilon}_{2m}$, respectively.
Applying the Arrhenius form, we obtain the relations
\begin{equation}
\begin{aligned}
q/\tilde{q} & = e^{-E_1/T_1} \,, \\
q_m/\tilde{q}_m & = e^{-E_m/T_1} \,, \\
q_{1m}/\tilde{q}_{1m} & = e^{-(E_1 - E_m)/T_1} \,, \\
\epsilon/\tilde{\epsilon} & = e^{-E_2/T_2} \,, \\
\epsilon_m/\tilde{\epsilon}_m & = e^{-E_m/T_2} \,, \\
\epsilon_{2m}/\tilde{\epsilon}_{2m} & = e^{-(E_2 - E_m)/T_2} \,.
\end{aligned}
\label{eq:three_level_relations}
\end{equation}

As in the two-level case, the net amount of work from the population difference in different energy levels (only for the second excited states in this case) is $\langle W_\mathrm{net} \rangle = (P_{1e} - P_{2e})(E_1 - E_2)$, where $P_{1e}$ and $P_{2e}$ refer to the population of $E_1$ in $R_1$ and $E_2$ in $R_2$, respectively.
The heat exchange, on the other hand, should take the $E_m$ level into account.
As a result, the efficiency for the three-level system is given by
\begin{equation}
\eta = \frac{(P_{1e} - P_{2e})(E_1 - E_2)}{(P_{1e} - P_{2e})E_1 + (P_{1m} - P_{2m})E_m} \,, 
\label{eq:three_level_efficiency}
\end{equation} 
where the term involving $E_m$, unless it vanishes, represents the extra heat exchange that cannot be used in the work extraction.

In contrast to the two-level case, the temporal part (involving $\tau_1$ and $\tau_2$) is not factorized in the functional form of $\langle W_\mathrm{net} \rangle / \tau = (P_{1e} - P_{2e})(E_1 - E_2)/\tau$ for this three-level case, so we cannot focus solely on the thermodynamics parameters. As shown in Fig.~\ref{fig:three_level_example}, 
the overall functional shape of power output $\langle W_\mathrm{net} \rangle / \tau$ varies over $\tau$ and the maximum value of power output occurs at different values of $(q^*,\epsilon^*)$ depending on $\tau$. Therefore, we conclude that the two-level system of our main interest is a special case that we can analyze deeply to obtain the insight presented so far. Moreover, for the three-level system, even at the limit $q \simeq \epsilon$ (which corresponds to the equilibrium or reversible limit for the two-level system, represented by the equilibrium distribution of population), there cannot be the equilibrium condition given by
\begin{equation}
\begin{aligned}
e^{-E_m/T_1} /Z & = e^{-E_m/T_2} /Z \,,\\
e^{-E_1/T_1} /Z & = e^{-E_2/T_2} /Z \,,
\end{aligned}
\label{eq:three_level_eq_population_condition}
\end{equation}
with the partition function $Z= 1+e^{-E_m/T_1} + e^{-E_1/T_1} = 1+ e^{-E_m/T_2} + e^{-E_2/T_2}$, unless $E_m = 0$. Hence, the condition of strong coupling between thermodynamic fluxes is also violated, which results in the linear coefficient of the efficiency $\eta_\mathrm{op}$ at the maximum power in terms of $\eta_C$ different from $1/2$~\cite{VanDenBroeck2005} when $E_m > 0$, and we have numerically verified the fact as well, where we have obtained $\eta_\mathrm{op}$ for the parameters $q$ and $\epsilon$, for given $E_m$ and $\tau$ values. In contrast, recall that for the two-level system, the strong coupling between thermodynamics always holds, and the reservoir symmetry approximately holds for $\eta_C \to 0$ (corresponding to $q^* \simeq \epsilon^*$ for the maximum power output). The necessity for the extra heat in Eq.~\eqref{eq:three_level_efficiency} also prevents the $\eta_\mathrm{op}$ from reaching unity at the other limiting case $\eta_C \to 1$, which we have also verified numerically.

\section{Conclusions and discussion}
\label{sec:conclusion}

We have demonstrated that our simple two-level heat engine model has a nontrivial parameter relation for the efficiency at the maximum power output. Thanks to the simplicity of our model composed of the two-level system, the time-dependent term only plays the role of an overall factor, so we have focused on the relative transition rates for a given temperature ratio of reservoirs. Based on numerical solutions and analytically driven asymptotic behaviors, we have shown that the optimal efficiency $\eta_\mathrm{op}$ for maximum power output in our model is clearly different from the Curzon-Ahlborn efficiency $\eta_\mathrm{CA}$~\cite{Chambadal1957,Novikov1958,Curzon1975}, although they share the same asymptotic behavior up to the quadratic term when $\eta_C \simeq 0$~\cite{VanDenBroeck2005,Esposito2009}. We have discussed its implication, in conjunction with the relation of the entropy production at the hot and cold reservoirs.

We have focused on the average thermodynamic quantities to yield the macroscopic efficiency in this paper, but it would be possible to consider the stochastic efficiency~\cite{Verley2014a,Verley2014b,Gingrich2014,Proesmans2015a,Proesmans2015b,Polettini2015} by adopting more specific protocols involved in the heat and work transfer, which can be a future work, along with the quantum effects~\cite{Scovil1959,Geusic1967,Bender2000,SAbe2011a,SAbe2011b,JWang2011,Uzdin2015,HJJeon2016,SLiu2016}. For comprehensive understanding, we would also need the full consideration of multi-level systems sketched in Sec.~\ref{sec:three_level} here, which we leave as future work.

\emph{Note added.}---Just before the submission of this paper, we have learned that a recent contribution by Toral \emph{et al}.~\cite{Toral2016} independently reports the same form of $\eta_\mathrm{op}$ in Ising spin systems or exclusion processes.

\begin{acknowledgments}
We thank Hyun-Myung Chun, Jae Dong Noh, Hee Joon Jeon, and Sang Wook Kim for fruitful discussions and comments.
\end{acknowledgments}

\end{document}